\input epsf.tex
\def\DESepsf(#1 width #2){\epsfxsize=#2 \epsfbox{#1}}
\documentstyle{aipproc}

\begin{document}

\rightline{OITS 635}
\rightline{SLAC-PUB-7601}

\title{BFKL Scattering at LEPII and a Next $e^+ e^-$ Collider}

\author{
S.\ J.\ Brodsky$^{*}$,
F.\ Hautmann$^{\dagger}$,
and D.\ E.\ Soper$^\dagger$
}

\address{
$^*$ Stanford Linear Accelerator Center, Stanford University,
Stanford, CA 94309\\
$^\dagger$ Institute of Theoretical Science, University of Oregon,
Eugene, OR 97403
}

\maketitle

\smallskip
\centerline{July 1997}
\bigskip

\leftline{Based on talks by D.\ E.\ Soper at DIS97
and by F.\ Hautmann at Photon97}

\begin{abstract}
We discuss virtual photon scattering in the region dominated by 
BFKL  exchange,   and report results for the cross sections at
present and  future 
$e^+ e^-$ colliders. 
\end{abstract}

The BFKL equation describes scattering processes in 
QCD in the 
limit of large energies  and fixed 
(sufficiently large) 
momentum transfers. 
The study that we present in this paper analyzes 
the prospects for using photon-photon collisions as a probe 
of QCD dynamics in this region. 
The quantity we focus on is   
 the total cross section for 
scattering 
two photons sufficiently far off shell at large 
center-of-mass energies, 
$ \gamma^* (Q_A^2) + \gamma^* (Q_B^2)  
\to {\mbox {hadrons}} $,  
$ s \gg Q_A^2 , Q_B^2 \gg \Lambda^2_{ QCD}$. 
This process can be observed at high-energy and high-luminosity 
$e^+ e^-$ colliders as well as $e^- e^-$ or $\mu^\pm \mu^-$ colliders, 
where the photons are produced from the lepton beams by bremsstrahlung. 
The $\gamma^* \gamma^*$ cross section can be measured in 
collisions 
in which both the outgoing leptons are tagged. 

The basic motivation  for this study is that 
compared to tests of BFKL dynamics in deeply inelastic lepton-hadron 
scattering (see, for instance, the review in Ref.~\cite{abra})  
  the off-shell photon cross section 
presents some theoretical advantages, essentially because it does not 
involve a non-perturbative target. The photons act as color dipoles 
with small transverse size, so that the QCD interactions can be 
treated in a fully perturbative framework. 

The structure of  
$\gamma^* \gamma^*$ high-energy scattering  
is shown schematically in Fig.~1.   
We work in a frame in which the 
photons $q_A, \, q_B$ have zero 
transverse momenta and are boosted along the positive and negative 
light-cone directions. 
In the leading logarithm approximation, 
the  process can be described  
as the interaction of two  $q {\bar q}$ pairs 
scattering off each other  
 via multiple gluon exchange. 
The $q {\bar q}$ pairs are in a color-singlet state and interact 
through their color dipole moments. The gluonic function ${\cal F}$ 
is obtained from the solution to the BFKL equation~\cite{bfkl}.

\begin{figure}
\centerline{\DESepsf(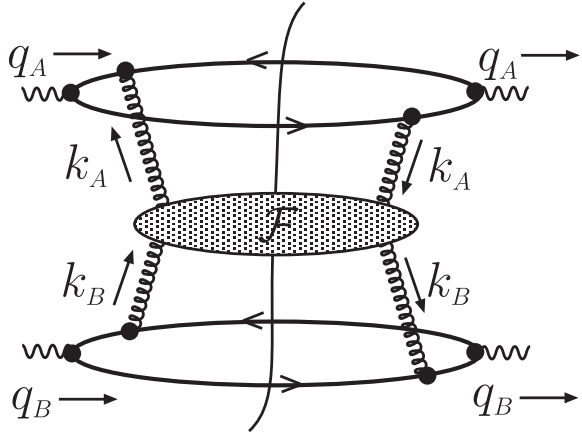 width 3 in)}
\caption{  The virtual photon cross section
in the high energy limit. 
\label{fac}}
\end{figure}

The analysis of the transverse-distance scales involved in 
the scattering illustrates a few distinctive features of this process.  
The mean transverse size of each $q {\bar q}$ 
dipole 
 is given, in the first 
approximation, by the reciprocal of the 
corresponding 
photon virtuality: 
\begin{equation}
\label{meanrt}
<R_{\perp \, A}> \, \sim 1 / Q_A \hspace*{0.4 cm} , \hspace*{0.6 cm}
<R_{\perp \, B}> \, \sim 1 / Q_B \hspace*{0.6 cm} .
\end{equation} 
 However,  
fluctuations can  
bring in 
much larger 
transverse sizes.  
Large-size fluctuations occur 
 as a result of the configurations in which  
one quark of the pair carries small 
transverse momentum and a small fraction of  the photon longitudinal 
momentum (the so-called 
aligned-jet configurations~\cite{aligned}). 
For example, for the  momentum $p_A$ of the quark created by photon 
$A$:
\begin{equation}
\label{alicfg}
{\bf p}_{\perp \, A} \ll  Q_A \hspace*{0.4 cm} , \hspace*{0.6 cm}
z_A \equiv p_A^{+}/q_A^{+} \ll 1  \hspace*{0.6 cm} .
\end{equation} 
The actual size up to which the $q {\bar q}$ pair can fluctuate 
is controlled by the 
scale of the system that it scatters off. Therefore, in 
  $\gamma^* \gamma^*$ 
scattering 
 the fluctuations  in the transverse 
size of each 
pair 
are suppressed by the 
off-shellness of the  photon creating the other pair. 
If {\em both} photons are sufficiently far off shell, the 
transverse separation in each $q {\bar q}$ dipole 
stays small~\cite{bhshep}.  
This can be contrasted with the case of 
deeply inelastic $e \, p$ 
scattering 
(or $e \, \gamma$, where $\gamma$ is a (quasi-)real 
photon). 
In this case, the  $q {\bar q}$ pair produced by the 
virtual photon can fluctuate up to sizes 
  of the order of a hadronic scale, that is, $1 / \Lambda_{QCD}$.  
This results in the 
 deeply inelastic  cross section 
being determined by an interplay of   
 short  and long distances.

In principle, the  $q {\bar q}$ dipoles in the 
$\gamma^* \gamma^*$ 
process  
 could  still 
 fluctuate 
to  bigger sizes  
in correspondence of 
configurations 
in which 
the jet alignment occurs twice, once for each photon. However, 
such configurations cost an extra overall power of $1 / Q^2$  
 in the cross section (terms proportional to  
$1 / (Q_A^2 \, Q_B^2 ) $ rather than 
$1 / (Q_A \, Q_B ) $)~\cite{bjfutu}. 
Therefore, they only contribute at the level of    
sub-leading power corrections  to 
$\sigma (\gamma^* \gamma^*)$. 

Even though the $q {\bar q}$ dipoles have small transverse size, 
sensitivity to large transverse distances may be brought in 
through the 
BFKL function   ${\cal F}$. This    indeed  
 is expected to  occur when the energy $s$ becomes very large. 
As $s$ increases, the typical impact parameters   
 dominating the cross section for BFKL exchange    
grow to be much larger than the size of the 
colliding objects~\cite{mdiff}. 
One can interpret this   as providing an upper bound on the range   
 of values of  $\left( \alpha_s (Q^2) \, \ln (s / Q^2) \right)$ 
in which the simple BFKL   
 approach to 
virtual photon  scattering 
 is expected 
to give reliable predictions~\cite{bhshep}.

The calculation of $\sigma (\gamma^* \gamma^*)$ 
and the 
form of the result  
 are discussed in detail in Refs.~\cite{bhshep,bhsprl}. 
 We recall here the main features:     

i) for large virtualities,  $\sigma (\gamma^* \gamma^*)$ scales like 
$1/Q^2$, where $Q^2 \sim {\mbox
{max}} \{ Q_A^2 , Q_B^2\} $. This is  characteristic of the 
perturbative QCD prediction. Models based 
on  Regge 
factorization 
(which work well in the soft-interaction regime  dominating 
$\gamma \, \gamma$ scattering near the mass shell) 
would predict a higher power in $1/Q$. 

ii) $\sigma (\gamma^* \gamma^*)$ is affected by  logarithmic corrections 
 in the energy $s$ 
to all orders in $\alpha_s$. As a result of the BFKL summation 
of these contributions, the cross section rises like   a power 
 in $s$, $\sigma \propto s^\lambda$. The Born  
approximation to this result (that is, the ${\cal O} (\alpha_s^2) $ 
contribution, corresponding to  single 
gluon exchange in the graph of 
Fig.~1) gives a constant cross section, $\sigma_{Born} \propto s^0$. 
This behavior in $s$ can be compared with lower-order 
calculations which do not include 
the  corrections associated to 
(single or multiple) 
gluon exchange. Such calculations 
would give cross sections that fall off like $1 / s$ at large $s$.

\begin{figure}[p]
\centerline{
\DESepsf(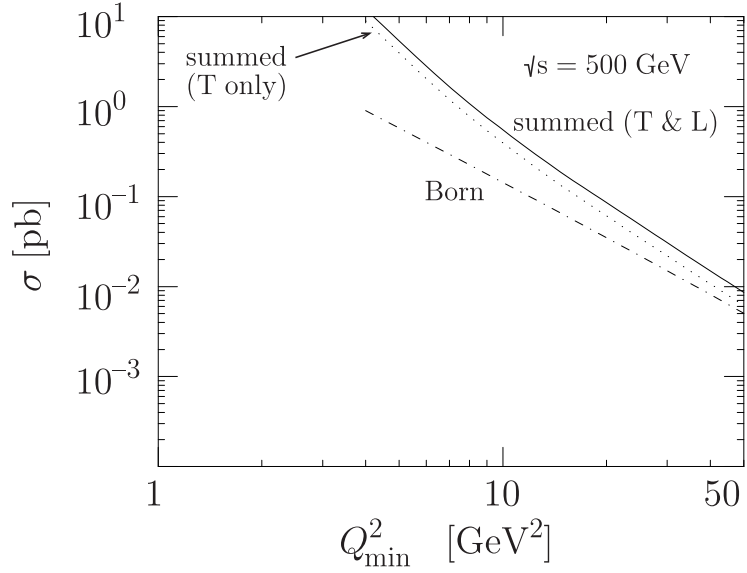 width 4.5 in)}
\caption{ The $Q_{\rm min}^2$ dependence of the $e^+ e^-$   integrated  
 rate
 for $\protect\sqrt s = 500\
{\rm GeV}$. The choice of the cuts and of the scales in the 
leading logarithm result is as in Ref.~\protect\cite{bhshep}. 
The dot-dashed and solid lines 
correspond to the result of using, respectively, the Born and the
BFKL-summed expressions for the photon-photon cross section. 
The dotted curve shows the contribution to the
 summed result coming from transversely polarized photons. 
\label{f500}}
\end{figure}

\begin{figure}[p]
\centerline{\DESepsf(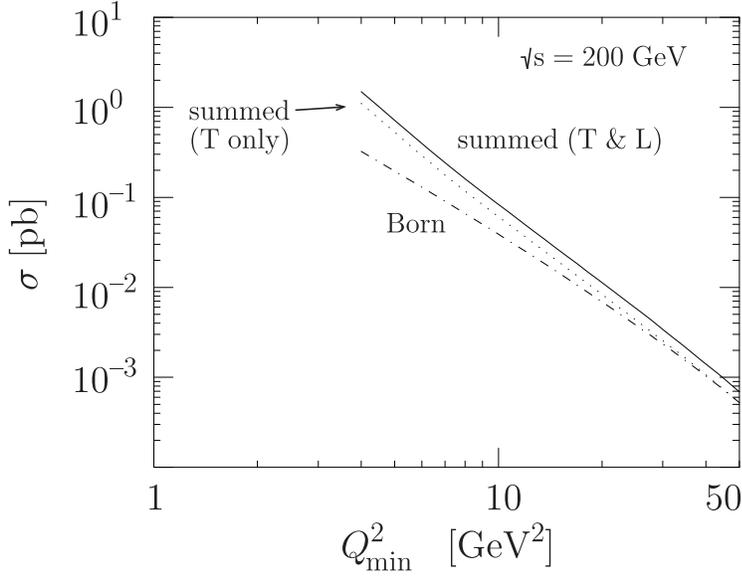 width 4.5 in)}
\caption{ Same as in Fig.~\protect\ref{f500} for $\protect\sqrt s =
200\ {\rm GeV}$. 
\label{f200}}
\end{figure}

These features are reflected 
 at the level of the  $e^+ e^-$ scattering
process.   
The $e^+ e^-$ cross section 
is obtained by folding  $ \sigma (\gamma^{*} \, \gamma^{*}) $ 
 with the flux of photons from each lepton.  
In Figs.~\ref{f500} and \ref{f200}, we  
 integrate  this cross section with a lower 
cut on the photon 
virtualities (in order that the coupling $\alpha_s$ be small, and that
the process be dominated by the perturbative contribution) 
and a lower cut on the photon-photon c.m.s. energy 
(in order that
the high energy approximation be valid). We plot the 
 result as a function of  the lower bound $Q_{\rm min}^2$, 
illustrating the 
expected dependence of 
the photon-photon cross section on the photon virtualities. 
  Fig.~\ref{f500} is for 
the energy of a future $e^+ e^-$ collider.  Fig.~\ref{f200}
 refers to the LEP collider  operating at
$\sqrt s = 200\ {\rm GeV}$. Details on our choice of cuts  
may be found in Ref.~\cite{bhshep}. 

From Figs.~\ref{f500} and \ref{f200}, for a value of the cut 
$Q_{\rm min} = 2 \ {\mbox {GeV}}$ we find
$ 
\sigma
\simeq 1.5 \, {\mbox {pb}} $ 
at LEP200 energies, and
$ 
\sigma
\simeq 12 \, {\mbox {pb}}
$ 
at the energy of a future collider. These cross sections would give
rise to about $750$ events at LEP200 for a value of the luminosity $L
= 500 \, {\mbox {pb}}^{-1}$, and about $6 \times 10^5$ events at
$\sqrt s = 500\ {\rm GeV}$ for $L = 50 \, {\mbox {fb}}^{-1}$.
 The above value of $Q_{\rm min}$ would imply 
 detecting leptons
scattered through angles down to about $20 \ {\mbox {mrad}}$ at
LEP200, and about $8\ {\mbox {mrad}}$ at a future $500 \, {\mbox {GeV}}$ 
collider.  If
instead we take, for instance, 
 $Q_{\rm min} = 6\  {\rm GeV}$, the minimum angle at a 
$500 \, {\mbox {GeV}}$ 
collider is
$24\ {\mbox {mrad}}$. Then the cross section is about $2 \times
10^{-2}\ {\rm pb}$, corresponding to about $10^3$ events.

\begin{figure}
\centerline{\DESepsf(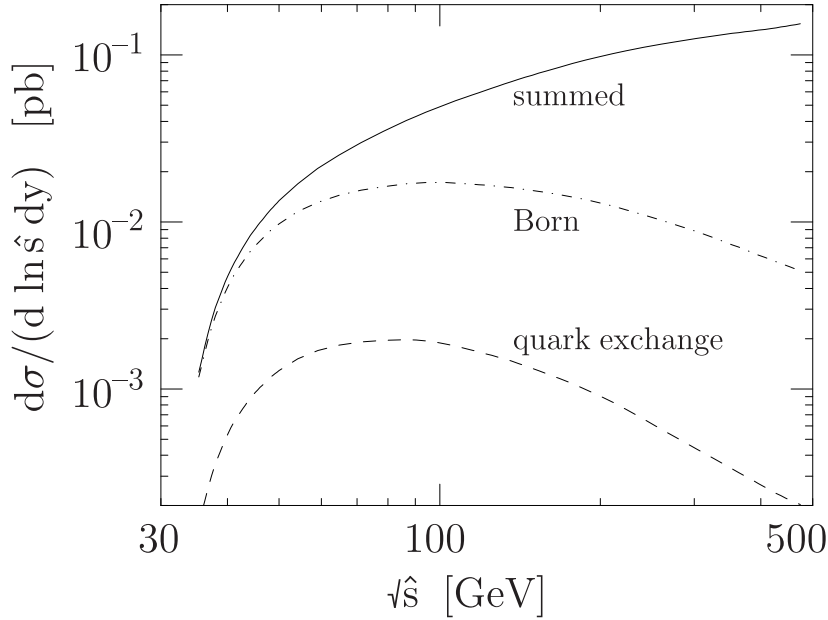 width 5 in)}
\caption{The cross section $d\sigma / (d \ln\hat s\, dy)$ 
 at $y = 0$ for $\protect\sqrt s = 500\
{\rm GeV}$. We take $Q_{\rm min}^2 = 10\ {\rm GeV}^2$.  
The solid curve is the summed BFKL result. The dot-dashed
curve is the Born result. The dashed curve shows the 
(purely electromagnetic) contribution 
arising from the scattering of (transversely polarized) photons via
quark exchange. 
\label{hats500}}
\end{figure}

The dependence on the photon-photon
c.m.\ energy $\sqrt {\hat s}$ can be best 
studied by fixing $Q_{\rm min}$ and  looking at the 
 cross section  
$d\sigma /( d \ln\hat s\, dy)$ (here $y$ is the 
photon-photon rapidity).  
 In Fig.~\ref{hats500} we plot this cross section 
at $y = 0$. 
While    
at the lowest end of the range in $\sqrt{\hat s}$  
the curves are 
strongly 
dependent    
on the choice of the cuts, for increasing $\sqrt{\hat s}$ the 
plotted distribution     
is rather directly related to the behavior of 
$\sigma (\gamma^* \gamma^*)$ 
discussed earlier. 
In particular, as $\sqrt{\hat s}$ increases to about $100 \
{\rm GeV}$ we see the Born result flatten out and the summed BFKL 
result rise, while the contribution from quark exchange is 
comparatively  suppressed.  The damping  towards the higher 
end of the range in $\sqrt{\hat s}$ affects  all curves and 
is due to the influence of   
  the photon flux factors.

Fig.~\ref{hats500} is for 
 $\sqrt{s} = 500\ {\rm
GeV}$. The corresponding curves 
at LEP200 energies 
are qualitatively 
similar. The main difference is that at $\sqrt{s} = 200\
{\rm GeV}$ there is less available range for $\sqrt{\hat s}$.

We see from the results presented above that at a future $e^+ e^-$
collider it should be possible to probe the effects of pomeron  
exchange in a range of $Q^2$ where summed perturbation theory
applies. One should be able to investigate this region in detail by
varying $Q_A$, $Q_B$ and ${\hat s}$
independently. At LEP200 such studies appear to be more problematic
mainly because of limitations in luminosity. Even with a modest
luminosity, however, one can access the region of relatively low
$Q^2$ 
 if one can get down to small
enough angles. This would allow one to examine experimentally the
transition between soft and hard scattering.

\medskip 
This work is supported in part by the United States Department of
Energy grants DE-AC03-76SF00515 and DE-FG03-96ER40969.


\begin{thebibliography}{99}
\bibitem{abra}
      H.\ Abramowicz, plenary talk at ICHEP96 (Warsaw, July 1996), 
      in Proceedings of the 
      XXVIII International  
      Conference on High Energy Physics,  
      eds. Z.\ Ajduk and A.K.\ Wroblewski, World Scientific, p.53.  
\bibitem{bfkl}
      L.N.\ Lipatov, 
          Sov.\ J.\ Nucl.\ Phys.\ {\bf 23}, 338 (1976); 
      E.A.\ Kuraev, L.N.\ Lipatov and V.S.\ Fadin, 
          Sov.\ Phys.\ JETP  {\bf 45}, 199 (1977) ;
      I.\ Balitskii and L.N.\ Lipatov,
          Sov.\ J.\ Nucl.\ Phys.\ {\bf 28}, 822 (1978).
\bibitem{aligned}
      J.D.\ Bjorken and J.\ Kogut, Phys.\ Rev.\ D {\bf 8}, 1341 (1973).
\bibitem{bhshep}
      S.J.\ Brodsky, F.\ Hautmann and D.E.\ Soper, preprint 
      OITS-629/97, SLAC-PUB-7480, e-print archive hep-ph/9706427.  
\bibitem{bjfutu}
      J.D.\ Bjorken, preprint SLAC-PUB-7341, presented 
      at Snowmass 1996 Summer Study 
      on New Directions for High Energy Physics,   
      e-print archive hep-ph/9610516. 
\bibitem{mdiff} 
      A.H.\ Mueller, Nucl.\ Phys.\ {\bf B437}, 107 (1995). 
\bibitem{bhsprl}
      S.J.\ Brodsky, F.\ Hautmann and D.E.\ Soper, Phys.\ Rev.\ Lett.\ 
      {\bf 78}, 803 (1997);  
      F.\ Hautmann,   talk at ICHEP96 (Warsaw, July 1996), 
      preprint OITS 613/96,    
      in Proceedings of the
       XXVIII International Conference on High Energy Physics,         
      eds. Z.\ Ajduk and A.K.\ Wroblewski, World Scientific, p.705;  
      J.\ Bartels, A.\ De Roeck and H.\ Lotter, 
      Phys.\ Lett.\ B {\bf 389}, 742 (1996).          

\end{thebibliography}
\end{document}